\documentclass[twocolumn,showpacs,preprintnumbers,amsmath,amssymb]{revtex4}
\usepackage{graphicx}
\usepackage{dcolumn}
\usepackage{epsfig}
\begin{document}
\title{\begin{flushright} \begin{small}
astro-ph/0209016
\end{small} \end{flushright}
Cooling Properties of Cloudy Bag Strange Stars}
\author{C.~Y.~Ng$^{1,3}$, K.~S.~Cheng$^1$, \& M.~C.~Chu$^2$ }
\affiliation{
$^1$Department of Physics, University of Hong Kong, Hong Kong, China\\
$^2$Department of Physics, The Chinese University of Hong Kong, Hong Kong, China\\
$^3$Department of Physics, Stanford University, USA\\
}
\date{\today}
\begin{abstract}
As the chiral symmetry is widely recognized as an important driver of the strong interaction dynamics, current strange stars models based on MIT bag models do not obey such symmetry. We investigate properties of bare strange stars using the Cloudy Bag Model, in which a pion cloud coupled to the quark-confining bag is introduced such that chiral symmetry is conserved. 
The parameters in the model, namely the bag constant and strange quark mass are determined self-consistently by fitting the mass spectrum of baryons. Then the equation of state is obtained by evaluating the energy-momentum tensor of the system. We find that the stellar properties of the Cloudy Bag Strange Stars are
similar to those of MIT Bag Models. However, the decay of pions is a very efficient cooling way. In fact it can carry out most the thermal energy in a few milliseconds and directly convert them into 100MeV photons via pion decay. This may be a very efficient $\gamma$-ray burst mechanism. Numerical results indicate that temperature of a Cloudy Bag strange star is sufficiently lower than a MIT one for the small gap energy of color superconductivity($\Delta$ = 1MeV). On the other hand, large gap energy ($\Delta$ = 100MeV) can suppress the pion emissivity and hence the cooling curves of Cloudy model and MIT model are almost identical. The long term cooling behaviors of both MIT model and Cloudy model are determined by the color-flavor locked phase. The surface luminosity of a bare strange star is higher than that of a neutron star until $10^6s$ and $10^8s$ for ($\Delta$ = 100MeV) and ($\Delta$ = 1MeV) respectively. After this period, the surface luminosity of a bare strange star becomes lower than that of a neutron star even rapidly cooling mechanisms, e.g. direct URCA process or pion condensation, exist in the neutron stars. Hence, the cooling behavior may provide a possible way to distinguish a compact object between a neutron star, MIT strange star and Cloudy Bag strange star in observations.
\end{abstract}
\maketitle

\section{Introduction}

It has been argued that strange quark matter, consisting of $u$-, $d$- 
and $s$-quarks, is energetically the most favorable state of quark matter~\citep{bo71, ta79, wi84}. \citet{wi84} suggested that there
are two ways to form strange quark matter: the quark-hadron phase
transition in the early universe and conversion of neutron stars into
strange ones at ultrahigh densities. If this strange matter
hypothesis is correct, then it has profound impact on physics and
astrophysics. Pulsars could be strange stars and not neutron stars as previously 
thought, and there could even be many strange dwarfs and strange planets in the
universe~\citep{al86, ha86, ma91, gl97, xu2001}.

Several mechanisms have been proposed for the formation of strange
quark stars. For example, strange stars are expected to form during the collapse of the 
core of a massive star after a supernova explosion~\citep{da95}. Another
possibility is that some rapidly spinning neutron stars in low-mass 
X-ray binaries (LMXBs) can accrete sufficient mass to initiate a phase 
transition to become strange stars~\citep{ch96}. Some of the millisecond pulsars
may be strange stars, because LMXBs are believed to be the progenitors 
of millisecond pulsars. Strange stars have also been proposed as 
sources of unusual astrophysical phenomena, such as soft 
$\gamma $-ray repeaters~\citep{ch98c, cd2002},
pulsating X-ray bursters~\citep{ch98b},
cosmological $\gamma $-ray bursts~\citep{ch96, da95, cheng2001},
SAX J1808.4-3658~\citep{li99} etc.
The discovery of kHz quasi-periodic
oscillations in LMXBs~\citep{zh98} implies that the compact stellar
object must have a very soft equation of state (EOS), which is
consistent with that of strange stars~\citep{kl98}. Recently, Hong et al. ~\citep{h2001}have argued that the first order phase transition of the color superconductivity occurring at the end point of some massive stars can release sufficient energy to produce hypernovae, which are considered as progenitors of cosmological $\gamma$-ray bursts.

However, important as this strange-matter hypothesis maybe,
it remains notoriously difficult to be proven or refuted on either
observational or theoretical grounds. The binding energy per nucleon
for strange matter is estimated to be close to that of Fe$^{56}$,
but it has not been possible to calculate it from QCD to within the
few percents level required to address the question. Model calculations 
of the binding energy are clearly inadequate and unreliable~\citep{fa84, ma91, gl97}.

Lacking an accurate and reliable method to calculate the strange
matter EOS, the structure and stability of strange stars have been
addressed~\citep{fa84, al86, ha86, ma91, gl97} typically in simplified models such as the MIT Bag Model~\citep{ch74}.
The quarks are treated as relativistic free particles confined in an
impenetrable bag. Approximately then, the pressure comprises of
the fermi gas pressure of the quarks minus a bag pressure, which mimics
the strong interaction that holds the quarks together. Perturbative
corrections to such a simple EOS can be made based on one-gluon exchange
calculations~\citep{fa84}. With this EOS, it has been shown that self-bound stars
exist with virtually no lower limit on the stellar mass~\citep{al86, ha86}; these
stars are basically giant hadrons, held together by strong interaction
rather than gravity.

It is arguable whether meaningful refinements of theoretical studies of
strange stars can be made at this stage. Uncertainties in the strong
interaction calculations easily make or kill strange stars, and 
arguments over details in model-dependent EOS's are probably futile. 
Nevertheless, one could ask what kinds of effects commonly accepted 
ingredients of strong interaction have on the stability and structure 
of strange stars. Whereas such studies are not meant to be 
quantitatively accurate, one 
may possibly gain insight into the physics of compact stars.

With this in mind, we study the implications of chiral symmetry on
the structure of strange stars. Chiral symmetry is widely recognized
as an important driver of the strong interaction dynamics, witness
the celebrated experimental success of partially conserved axial
current (PCAC), which signifies the importance of chiral SU(2)xSU(2)
symmetry~\citep{bh88}. Phenomenologically, the partial breaking of chiral
symmetry is associated with the emergence of pion as a pseudo-Goldstone
boson and hence related to the small pion mass~\citep{bh88}; the coupling of 
pions to quarks/hadrons is thus a natural result of chiral dynamics 
and an essential ingredient of strong interaction physics.

There are many hadron models that incorporate various features of QCD 
~\citep{bh88}. Whereas the MIT Bag Model has the asymptotic freedom and the
confinement of quarks built in, it is well known to violate
chiral symmetry badly.
By coupling the pions to the quarks, thus savaging the axial current at
the bag, one can restore chiral symmetry. There are many variants of
such chiral bag models~\citep{my84}, and one can imagine these as either 
modifying the original bag model to include a meson cloud at the bag, 
or extending the sigma model to a nucleon with quark structure.
For our purpose, a simple, representative chiral bag model suffices, and
we have thus chosen the Cloudy Bag model~\citep{th83, mi84}.

\section{Cloudy Bag Model Equation of State}

The idea to model a hadron by putting a few relativistic quarks in a 
cavity has led to much phenomenological success~\citep{ch74, bh88}. In the simplest 
version, one considers free Dirac fermions of mass $m_q$ confined 
in a static spherical bag of radius $R$, and the ground state 
spatial wavefunction is simply ~\citep{bh88, th83}
\begin{equation}
q(r) = {N\over \sqrt{4\pi}} \left[ gj_0(xr/R), 0, ih{\bf \sigma}\cdot
{\bf \hat r} j_1(xr/R), 0 \right] \ \ ,
\end{equation}
where $x\equiv R\sqrt{E^2-m_q^2}$, $g\equiv \sqrt{(E + m_q)/E}$,
$h \equiv \sqrt{(E - m_q)/E}$,
$N\equiv \sqrt{{E(E-m_q)\over[2E(E-1/R)+m_q/R]R^3j_0^2(x)}}$ 
and $x$ satisfies
$\tan x = {x \over{1-m_qR - \sqrt{[x^2+(m_qR)^2]} } }$.

However, it was soon realized that the original MIT bag boundary 
condition leads to severe violation of chiral symmetry~\citep{bh88, th83}.
The physical picture is that when a quark with a certain helicity 
collides with the impenetrable wall, its momentum is reversed 
but not its spin, leading to a change of the helicity. 
Mathematically, this is expressed by the
non-conservation of the quark axial current ${\bf A^q} = {\bar q} 
\gamma _5 {\bf \gamma \tau}q/2$, where the $\gamma$'s are the 
Dirac gamma matrices and ${\bf \tau}$ is the isospin operator. 
One resolution of the problem is to
introduce a meson field, so that the total (quark plus meson) axial 
current is conserved~\citep{bh88, th83}. Physically, when a quark hits the bag, 
it produces mesons that carry away some of the axial current. 
In this line of thought,
the hadron bag is always surrounded by a pion cloud, hence the
name Cloudy Bag Model. For example, for a ground state nucleon, the 
pion field that will restore the chiral symmetry of a bare MIT bag is
\begin{equation}
{\bf \phi} = \phi (r) ({\bf \sigma} \cdot {\bf \hat r} ) {\bf \tau} \ \ ,
\end{equation}
where ${\bf \sigma}$ is the Pauli matrix, and the spatial part of the
wavefunction $\phi(r)$ is given by
\begin{equation}
\phi = a {1+\mu r \over r^2} e^{-\mu r} \ \ , \ r> R \ \ ,
\end{equation}
Here the pion-quark coupling constant $a$ is determined by the continunity of axial current at $r=R$, which yields
\begin{equation}
a=-\left(\frac{E^2-m^2}{2E(E-\frac{1}{R})+\frac{m}{R}}\right)\left(\frac{n}{4\pi f}\right)\left(\frac{1}{2(1+\bar \mu)+\bar \mu^2}\right) e^{\bar \mu} \ \ ,
\end{equation}
with $n$ is the total number of $u$ and $d$ quarks in the bag, $\mu$ is the meson mass, $f$ is the pion decay constant and
${\bar \mu} \equiv \mu R$.

The Cloudy Bag Model Lagrangian is a simple extension of that of the
MIT Bag Model. Here we use the linearized version:
\begin{widetext}
\begin{eqnarray}
{\cal L}_{\rm CBM} &= &\left[ {i \over 2} \left({\bar q}(x) \gamma ^{\mu}
\partial_{\mu} q - \partial _ {\mu} {\bar q} \gamma ^{\mu} q \right)
- m_q {\bar q} q - B \right] \theta _V \nonumber \\
&&- {1 \over 2} {\bar q} \left( 1 + i{\bf \tau} \cdot
{\bf \phi} \gamma _5 /f \right) q \Delta _S + {1\over 2} \left[ \partial _{\mu}
{\bf \phi} (x) \cdot \partial ^ {\mu} {\bf \phi} (x) - \mu ^2 
{\bf \phi}^2 \right]\bar \theta_V \ \ ,
\end{eqnarray}
\end{widetext}
where $\theta _V = 1 (0)$ inside (outside) the bag, $\bar \theta_V=1-\theta_V$, $\Delta _S $
is the surface delta function, and $B$ is the bag constant.
Note that the quarks are completely confined in the bag, and the
mesons only exist outside the bag. They are free particles andare coupled to the quarks only at the bag surface, through
the term proportional to $\Delta _S$, and are free outside or inside the
bag. The Chiral Bag Model has essentially the same Langrangian but 
differs from this one in that it excludes the mesons from the 
inside of the bag~\citep{br79, bh88, th83}.

>From the Lagrangian, one can obtain the stress-energy tensor:
\\$T^{\mu \nu} = -g^{\mu \nu} {\cal L} + \left[ {\partial {\cal L} \over
\partial (\partial _{\mu} q) } \partial ^{\nu} q + \partial ^{\nu} 
{\bar q}
{\partial {\cal L} \over \partial (\partial _{\mu} {\bar q} )} \right]$,
which gives the pressure and density of the system immediately,
$P = {1 \over 3} T_{ii}$, $\rho = T_{00}$.
With some straightforward algebra, we get:
\begin{equation}
P_q =  \left\{ {N^2 \over 12 \pi} {E^2 - m_q^2 \over E} 
\left[ j_0^2
\left({xr \over R}\right) +j_1^2 \left({xr \over R}\right) \right] -B
\right\} \theta _V \ \ ,
\label{quarkp}
\end{equation}
\begin{equation}
\rho _q = \left\{  {N^2 \over 4 \pi} \left[ (E+m_q)j_0^2
\left({xr \over R}\right) +(E-m)j_1^2 \left({xr \over R}\right) 
\right] +B \right\} \theta _V
\label{quarkrho}
\end{equation}
inside the bag, Note that if one ignores the quark mass, the EOS becomes
\begin{equation}
P = \sum_q P_q =\frac{1}{3}\left( \sum_q \rho_q - 4B\right) \ \ .
\end{equation}
which is exactly the one used in previous studies of strange stars~\citep{al86, ha86}.

We assume that for a macroscopic object like a strange star, the pressure outside the bag due to the meson cloud
is dominated by real and free pions, given by 
\begin{equation}
P_{\phi} ={1 \over 3} \rho_{\phi}c^2 \quad .
\label{peos}
\end{equation}
At the boundary, continunity of pressure is required. Therefore, when the quark pressure equals to the pion pressure just outside the bag, the EOS is truncated to the pion one eq.~\ref{peos}.

In order to deduce the EOS and apply it to study the properties
of strange star, we need to determine the values of the bag constant ($B$) and
the strange quark mass ($m_s$). For simplicity we have taken the masses of up and down quarks
to be zero. First, the hadron mass can be expressed as
\begin{equation}
E(R) = \frac{4\pi}{3} B R^3 + \sum_i \frac{x_i}{R} + \int \rho_{\phi}
d^3r \ \ ,
\end{equation}
where $R$ is the radius of the hadron, $x_i$ is the eigenvalue of quarks
with $i=$ u, d and s.
$R$ is fixed by minimizing the mass, i.e., $\frac{\partial E(R)}{\partial R} = 0$. 
Then the hadron mass can be determined. Since $B$ and $m_s$ are independent parameters, we first determine $B$ by fitting mass of the baryons which consist of only up and down quarks. In this simple model, hadrons with spin 1/2 and 3/2 are degenerate,
and we minimizing the averaged root-mean-square error of  $N$ and $\Delta$. Table~\ref{table:b} is the result of our fitting.
We obtain $B^{1/4} = 120\,$MeV.
After $B$ is fixed, $m_s$ can be determined by fitting baryons which consist of one or more strange quarks.
>From table.~\ref{table:ms}, we obtain $m_s$ = 220 MeV. 

\begin{figure*}[h]
\includegraphics[width=15cm]{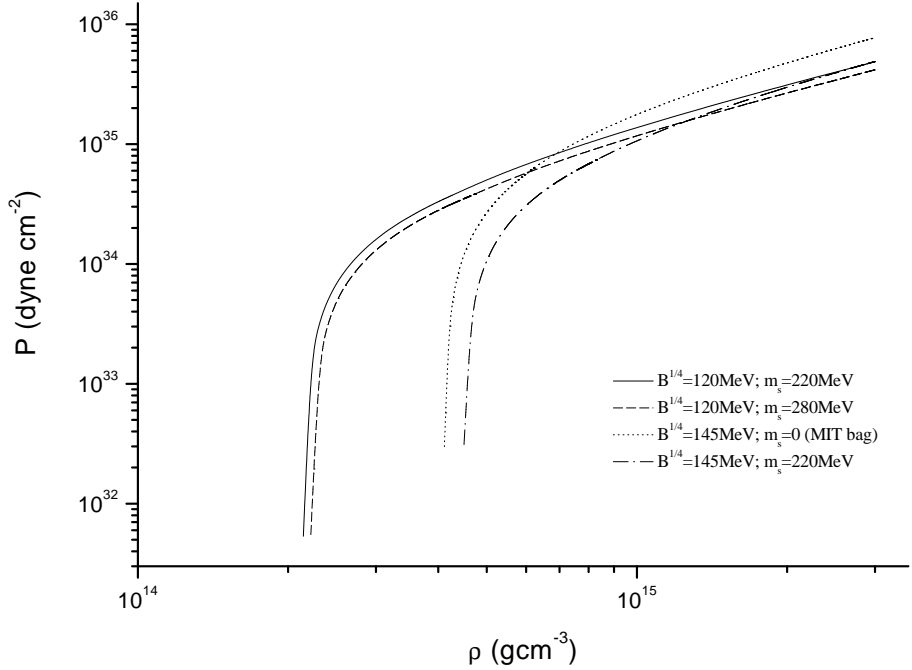}
\caption{EOS's for different parameters.}
\label{fig:eos}
\end{figure*}

Fig.~\ref{fig:eos} shows the EOS from the cloudy bag model. 
We have extrapolated it into high-density region.
In low-density region, since the pion field decay exponentially outside the bag, the pion `tail' in the EOS is not important in calculation. Hence, it is not shown in the graph. The EOS's for different parameters were also shown in the figure as for comparison.
Note that pion does not come into account in this part of the EOS, therefore the one with $B^{1/4}=145\,$MeV and $m_s=0\,$MeV reduced to the MIT bag case, which was used in previous studies of strange star.

\section{Stellar Properties of Strange Stars}
\subsection{Global Stellar Structure}
Assuming the pion cloud decay exponentially when leaving the stellar surface, it only exists within an effective range of scale $1/\mu$ ($\sim1$\,fm). The pion `shell' is so thin that has negligible effect on the stellar radius. Using typical parameters: $R\sim10^6$\,cm, $\Delta r\sim1$\,fm, \mbox{$\rho_\pi\sim10^{11}\,\mathrm{g\,cm^{-3}}$}, mass of the pion cloud surrounding the star can be estimated:
\[4\pi R^2\Delta r\rho_\pi\sim 10^{11}\,\mathrm{g}\sim 10^{-22}\,M_\odot \]
which is completely negligible as compared to the stellar mass. The pion cloud surrounding the star is a consequence of chiral symmetry consideration, it has no significance on global stellar structure such as radius and mass.

\begin{figure*}[h]
\includegraphics[width=15cm]{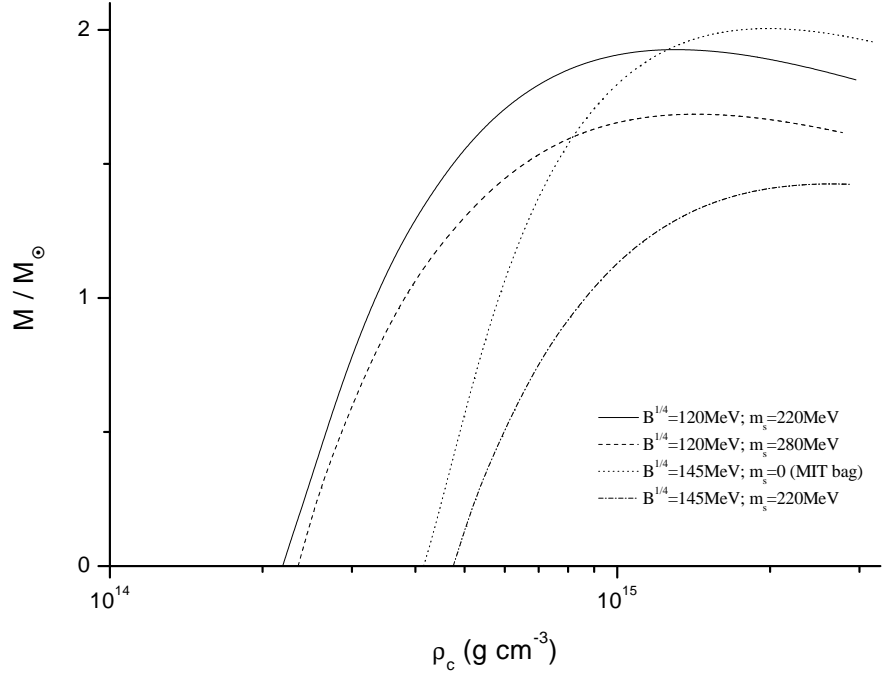}
\caption{Gravitational mass ($M$) versus central density ($\rho_c$) for
different parameters.}
\label{fig:mrhoc}
\end{figure*}

\begin{figure*}[h]
\includegraphics[width=15cm]{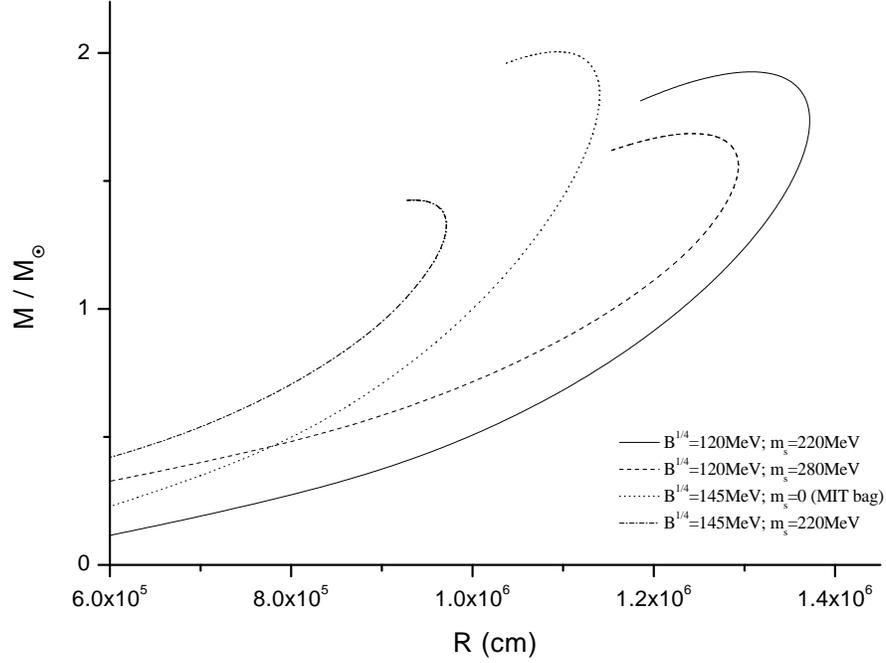}
\caption{Gravitational mass ($M$) versus stellar radius ($R$) for
different parameters.}
\label{fig:mr}
\end{figure*}

\begin{figure*}[h]
\includegraphics[width=15cm]{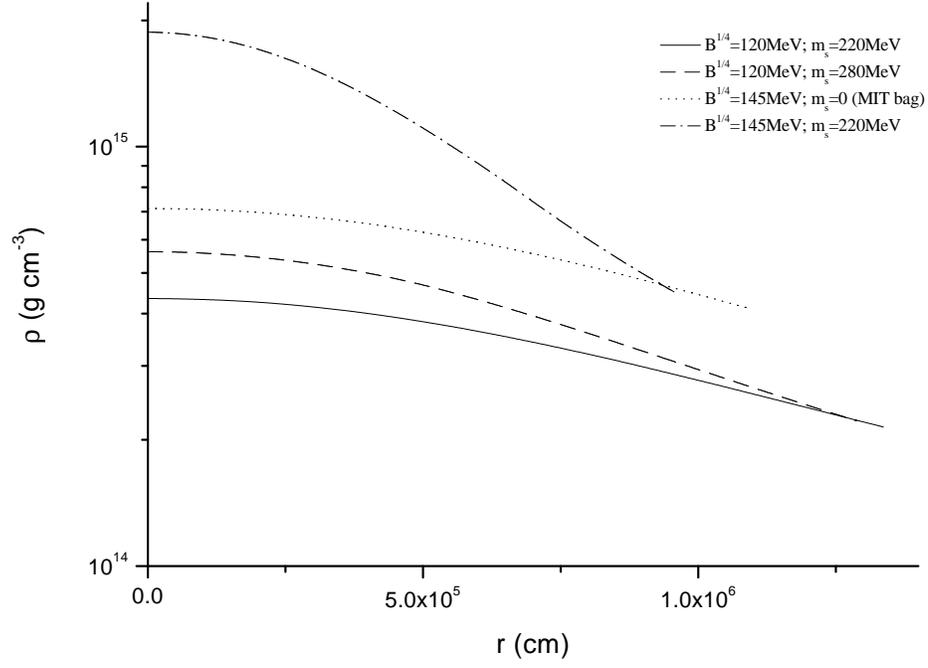}
\caption{The local density ($\rho$) at radial distance $r$ from the center
for $1.4\,M_\odot$ strange stars.}
\label{fig:rhor14}
\end{figure*}

Using the EOS of this model and the Tolman-Oppenheimer-Volkoff
equations, the properties of strange stars can be obtained. The
global stellar structure is very similar to that of MIT models~\citep{al86, ha86}.
>From fig.~\ref{fig:mrhoc}, maximum mass of Cloudy Bag Strange stars using
the best-fit parameters is about$1.9\,M_\odot$. As a comparison, the
maximum mass of MIT strange stars is 2.0\,$M_\odot$. These two values only differ slightly, $\sim 4 \%$. Similar to the case of MIT strange stars, there is no minimum mass for the Cloudy Bag strange stars. The mass-radius relation is shown in fig.~\ref{fig:mr}. From the figure, radius of the Cloudy Bag strange star is in general larger, this is due to a softer EOS obtained by a smaller value of $B$. The stellar density profile of $1.4\,M_\odot$ strange stars is shown in fig.~\ref{fig:rhor14}, the density is quite uniform inside the star.

\begin{figure*}[h]
\includegraphics[width=15cm]{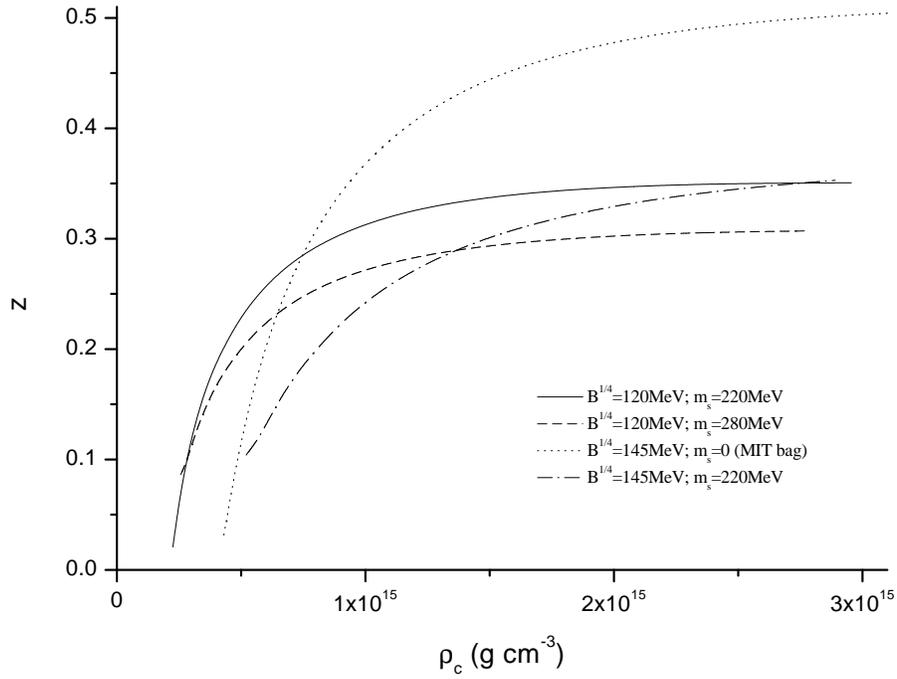}
\caption{Gravitational redshift ($z$) versus central density ($\rho_c$)
for different parameters.}
\label{fig:z}
\end{figure*}

Fig.~\ref{fig:z} is the gravitational redshift ($z$) versus central
density ($\rho_c$) for different parameters, where $z$ is given by
$z=\left(1-{2M\over R}\right)^{-1/2}-1$. From the figure, redshift depends
sensitively on strange quark mass $m_s$. This result can be understood
qualitatively from~\citet{wi84}: since both the mass and radius of the
star
depend
on the same factor of $B^{-1/2}$, their ratio, $M/R$ will be independent of $B$. Then the only parameter remaining is $m_s$. This may provide a possible way to estimate the strange quark mass $m_s$. If the gravitational redshift of strange stars is observed, $m_s$ may be inferred from the graph.

\subsection{Rotation Properties}
Rotation properties of strange star is important in observation, since the spin rate of pulsars can be measured very accurately. Due to different visocity of strange matter and neutron matter, the maximum rotation frequency of strange stars could be much higher than neutron stars~\citep{ma92}.
\citet{ch00} derived the approximate mass and radius formulae for static and rotating strange stars using the general relativistic conditions of thermodynamic equilibrium and fitting with values obtained from numerical integration of structure equations. From the formulae, the maximum mass and radius of strange stars can be predicted by fitting only one set of static configuration data.

In our calculation, we will adopt the formulae. The equatorial radius of a rotating strange star can be expressed as a function of static mass ($M_\mathrm{stat}$) and static radius ($R_\mathrm{stat}$):
\begin{equation}
R_e\approx \left(\frac{M_\mathrm{stat}}{R_\mathrm{stat}}\right )^{1/2}\left[ \frac{M_\mathrm{stat}}{R_\mathrm{stat}^3}-\frac{9\Omega^2}{4G}\left(1-\frac{2GM_\mathrm{stat}}{c^2R_\mathrm{stat}}\right)\right]^{-1/2},
\end{equation}
where $G$ is the constant of gravitation and $\Omega$ is the angular velocity of rotation.
Gravitational mass of a rotating strange is assumed to be in the form
\begin{equation}
M_\mathrm{rot}\approx M_\mathrm{stat}\left( \frac{R_e}{R_\mathrm{stat}}\right)^3e,
\end{equation}
where $e$ is the eccentricity, defined by the ratio between polar and
equatorial radii, $R_p/R_e$. We used the approximation as given by~\citet{wb92}:
\begin{equation}
e\approx \frac{2R_\mathrm{stat}}{R_e}-1.
\end{equation}
Now the gravitational mass is a function of rotation frequency. The maximum frequency is given by the stability condition
\begin{equation}
\frac{\partial M}{\partial \Omega}=0.
\end{equation}

\begin{figure*}[h]
\includegraphics[width=15cm]{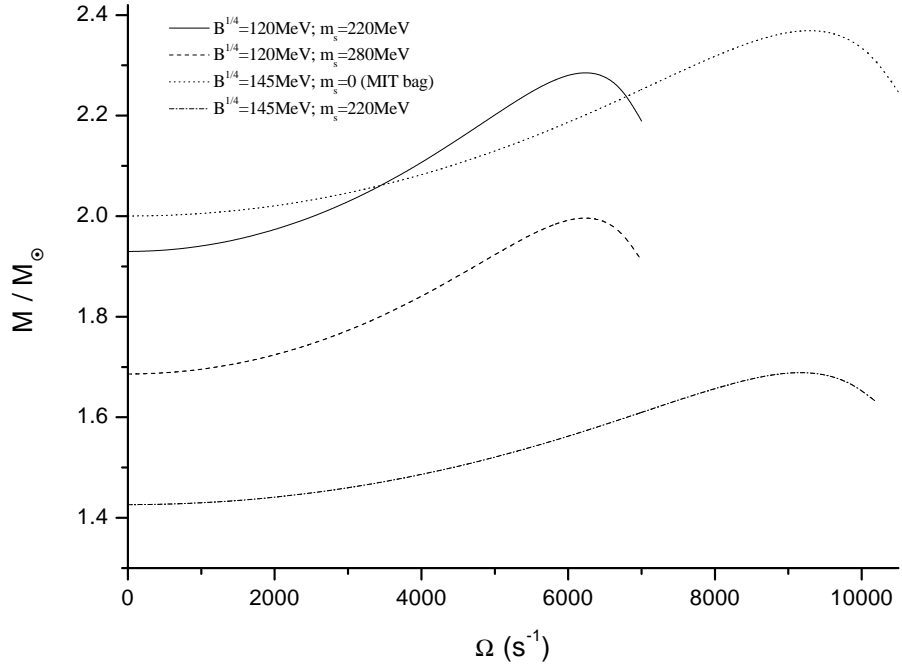}
\caption{Gravitational mass ($M$) as a function of rotation frequency
($\Omega$) for stars with maximum masses.}
\label{fig:mw}
\end{figure*}

Fig.~\ref{fig:mw} shows the gravitational mass ($M$) versus the rotation frequency ($\Omega$) for different strange stars with maximum masses. According to the graph, the maximum rotation frequency of the Cloudy Bag strange stars with best-fit parameters is 6300\,s$^{-1}$, which corresponds to the maximum mass of $2.3\,M_\odot$. For MIT strange stars, the Kepler limit is about 9400\,s$^{-1}$ and the maximum mass is $2.4\,M_\odot$.
The maximum rotation frequency of a MIT strange star is higher than the Cloudy Bag one by $\sim$30\%. This is due to different mass-radius properties. A MIT strange star in general more compact, having a smaller radius and larger mass. Hence the rotation can be faster.

\begin{figure*}[h]
\includegraphics[width=15cm]{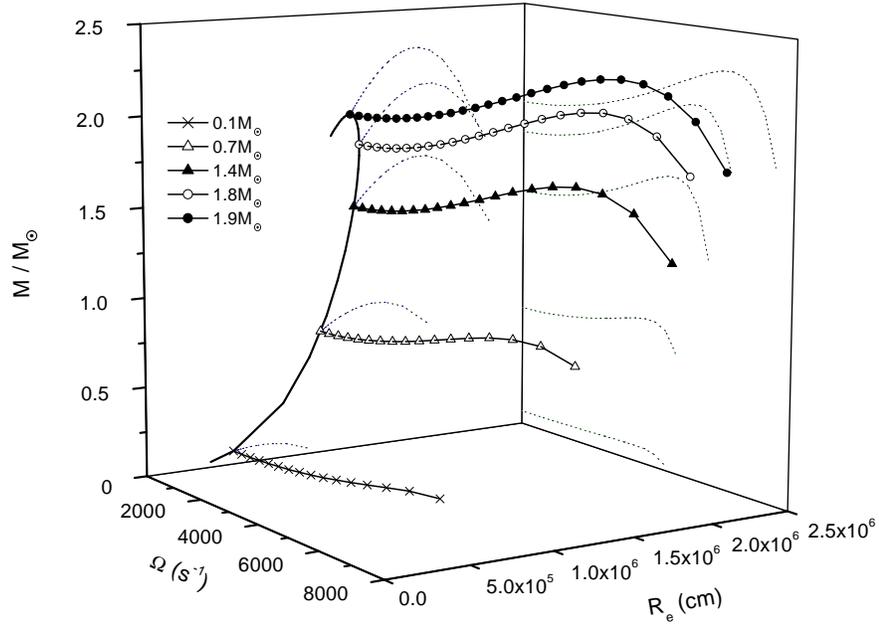}
\caption{Gravitational mass ($M$) versus equatorial radius ($R_e$) and
rotation frequency ($\Omega$) for Cloudy Bag strange stars with best-fit
parameters.}
\label{fig:rotat}
\end{figure*}

In order to have a better understanding on the evolution of gravitational mass and equatorial radius as the spin rate increases, fig.~\ref{fig:rotat} is the plot of $M$ against $R_e$ and $\Omega$ for the best-fit Cloudy Bag strange stars with different static masses. The thick solid line on the $M$-$R_e$ plane is the mass-radius relation of the static stars, i.e. fig.~\ref{fig:mr}. Then stars with static masses of $0.1\,M_\odot$ $0.7\,M_\odot$, $1.4\,M_\odot$, $1.8\,M_\odot$ and $1.9\,M_\odot$ were investigated as rotation frequency increases.

>From the graph, $R_e$ increases monotonically with $\Omega$, while the mass attends a maximum value and then drops with $\Omega$. The change in mass is not so obvious for a low mass strange star, e.g. $0.1\,M_\odot$. For the case of a $1.4\,M_\odot$ star, the maximum mass is $\sim 1.7\,M_\odot$ which corresponds to the Kepler frequency of 5000\,s$^{-1}$ and the equatorial radius of $1.8\times10^6$\,cm. The change in mass is about 20\% and in radius is about 40\%. For the star with maximum static mass, the Kepler limit is about 6300\,s$^{-1}$, with $M=2.3\,M_\odot$ and $R_e=1.8\times 10^{6}$\,cm. The evolution of mass and radius can be investigated by the projection of the curves on the $M$-$R_e$ plane. They are shown by the dotted lines. The gravitational mass as well as equatorial radius increase with rotation frequency, until the maximum mass is reached.

Projection of the curves on the $\Omega$-$M$ plane is also shown by the dotted lines. They demonstrate clearly the relationship between maximum rotation frequency and static stellar mass. From the graph, the Kepler frequency increases with static stellar mass. Hence the maximum possible mass of a stable star is the one with largest static mass and rotating at Kepler frequency. The graph indicates that if a strange star with period less than 1\,ms is found, it probably has a mass greater than 1.4\,$M_\odot$.

Table~\ref{table:summary} gives a summary of the results for different parameters. For a rotating strange star, the mass and equatorial radius lie in different region from the static case. Using the formulae, for given any two stellar parameters from the gravitational mass, equatorial radius or rotation frequency, the remaining one can be determined from the graph. This may help to determine the size or mass of a strange star since rotating period of a pulsar can be measure very accurately. 

\section{Cooling Properties of Cloudy Bag Strange Stars}
Since the proposal of strange stars, much effort has been devoted to find

observational properties that distinguish strange stars from neutron
stars. Although many stellar properties are similar in both kinds of
stars, their cooling behaviour is very different. Energy lost rate of
strange matter is much higher than that of neutron matter, therefore the
temperature of a young strange star is sufficiently lower than a neutron
star of the same age~\citep{pi91, ch96b, sc96}.

By chiral symmetry consideration, we suggested that a strange star should be surrounded by a pion field. The thin pion cloud has negligible effect on global stellar structure such as mass and radius. However, it plays an important part in strange star cooling. The decay of pion is a rapid cooling mechanism in our model. As a result, the surface temperature of a Cloudy Bag strange star is lower than ordinary MIT strange star significantly. This may provide a hint to distinguish Cloudy Bag strange stars and verify the theory. In addition, when the star is cooled by emission of pions, the huge amount of energy released within a short period may be a possible energy source of $\gamma$-ray bursts.

\subsection{Assumptions}
In this section, we considered the cooling of a non-rotating bare strange star with typical mass 1.4$M_\odot$, which is the typical mass of neutron star. 
Indeed, there is no reason why a strange star and a neutron star should have the same masses. Also it is unnecessary that there are identical twin between strange star and neutron star. However, the radius of the strange star is insensitive to the stellar mass. For example, the radius variation of strange star with masses between 1.4 to 2 solar masses is only 3$\%$. Furthermore, if the internal properties of the strange stars are the same, the cooling curves are almost identical. Of course, the density profiles of strange stars with different mass are different. As we will point out in next section, strange stars with different masses will have different electron density due to the color-flavor locked (CFL) phase ~\citep{r2001}. 
This has a very important effect in cooling calculations.
Since our focus in this paper is on the role of pion cloud surrounding a strange quark star, which is a natural consequence of chiral symmetry, and in particular, its effect on the cooling properties on strange stars. We are not primarily interested in the detail of other effects, i.e. CFL of color superconductivity, in this paper. Therefore, in this paper we treat electron density as a parameter but the mass of the strange star is fixed as $1.4 \,M_\odot$ in the cooling calculations. For simplicity we assume that all compact star models are non-rotating. As argued by~\citet{us98}, when a
strange star
is formed, it is very hot and most of the normal matter in the curst is ejected. Therefore, it is expected that new born strange stars do not have crusts. Besides, we assumed that all pions will decay once they leave the stellar surface beyond the length scale of $1/\mu$.

For simplicity, we assumed the star of uniform density and isothermal. >From figure~\ref{fig:rhor14}, the uniform density profile is a good approximation. Also, the isothermal approximation is generally used in simplified models. As it is a bare strange star in our case, the core and surface temperature are equal. Note that the effects of magnetic field are neglected in this simple model. This will be carried out in further works.

Superfluidity (superconductivity) of quark matter is considered in the model. It is
generally believed that quarks can form colored cooper pairs near the Fermi surface and
become superfluid (superconducting). An early estimation of the magnitude of the gap
$\Delta$ was $\sim $\,0.1-1~MeV~\citep{ba84}. On the other hand, some recent studies
considered instanton-induced interaction between quarks and estimated $\Delta\sim
100$\,MeV~\citep{al98, al99a, al99b, ra98, ra99, sc98, ca99, bl98, pi00, sc99a, sc99b}.
Furthermore, recent works (\citep{al98, al99b, ra98, ra99, sc98, ca99, bl98, bl99,
pi00, al99a, sc99a, sc99b}) demonstrate
the possibilities of diquark condensates characterized by large pairing gaps in quark cores
of some neutron stars and in quark stars and discuss different possible phases of quark
matter, for examples, the color-phase-locked u-d-s phase (\citep{al99a, al99b, sc99a,
sc99b}) and two flavor u-d color-superconducting phase (\citep{al98, ra98, sc98, bl98, 
ca99, bl99, pi00}). In this paper, our main focus
is on the role of the
pion cloud surrounding a quark star and its effect on the cooling properties of strange stars. We are not primarily interested in the detailed effects of color superconductivity, which has been discussed in
some details in Ref. (\citet{b2000}). Furthermore, there are large uncertainties as to the
phase as well
as the parameter values of color superconductivity in quark stars. Therefore we will assume
that there are three flavors of u-d-s quarks in a strange star for both small and large
color superconductivity gaps. Another main difference of these phases is the electron
concentration, which plays a very important role in cooling calculations and will be
discussed later. From BCS theory, the critical temperature $T_c$ is related to the pairing
gap by $k_BT_c=\Delta/1.76$ where $k_B$ is the Boltzmann constant. When the temperature is
below $T_c$, it is in the superfluid state and there are some modifications to the specific
heat and emissivity.

\subsection{Cooling Processes}
In the cooling of Cloudy Bag strange stars, the main cooling processes include quark URCA process, pion emission and blackbody radiation. The thermal evolution of the system is determined by the equation:
\begin{equation}
C\frac{dT}{dt}=-(L_\nu+L_\pi+L_{bb}),
\end{equation}
where $C$ is the specific heat, $L_\nu$, $L_\pi$, $L_{bb}$ are neutrino emissivity, pion emissivity and blackbody radiation respectively.

In normal state, we adopted the results of~\citep{iw80},
where
the quark Fermi momentum was approximated by:
\begin{equation}
p_F(q)=235\left( \rho/\rho_0\right) ^{1/3}\mathrm{MeV}/c,
\label{eq:pf}
\end{equation}
the specific heat of quark matter is then given by:
\begin{equation}
c_q=2.5\times 10^{20}\left( \rho/\rho_0\right)^{2/3}T_9\,\mathrm{erg\,cm^{-3}\,K^{-1}}.
\label{eq:cq}
\end{equation}
In superfluid state, the specific heat is modified, we adapted the formula
by~\citet{ho91a}
and~\citet{ma79}:
\begin{widetext}
\begin{equation}
\left. \begin{array}{lll}
c_q^{\mathrm{sf}} & = & 3.15c_q\left( \frac{T_c}{T}\right) e^{-1.76T_c/T}\left[\,2.5 -1.66\left(\frac{T}{T_c}\right) +3.64\left(\frac{T}{T_c}\right)^2\right] \\
&& \mathrm{\;for} \quad 0.2\,T_c\leq T\leq T_c \\
&=& 0 \quad \mathrm{\;for} \quad T< 0.2\,T_c.
\end{array} \right\}	\label{eq:csf}
\end{equation}
\end{widetext}
It has been shown that the Goldstone excitations in the quark-gluon plasma contribute to the specific heat of the strange star (~\citep{b2000},~\citep{b2001}). 
\begin{equation}
c_{g-{\gamma}}=3.0\times 10^{13} N_{ g-{\gamma}} T_9^3\,\mathrm{erg\,cm^{-3}\,K^{-1}},
\label{eq:gg}
\end{equation}
where $N_{g-{\gamma}}$ is the number of available massless gluon-photon states, which are present even in the color superconducting phase.
When temperature is low, the specific heat of quark matter vanishes. However, electrons are not affected. Therefore, the specific heat of electrons becomes important. We also considered this factor
\begin{eqnarray}
c_e &=& \frac{3^{1/3}}{3}\,2.5\times 10^{20}\left(\frac{Y_e\rho}{\rho_0}\right)^{2/3}T_9\,\mathrm{erg\,cm^{-3}\,K^{-1}}	\nonumber	\\
&=&1.7\times 10^{20}\left(\frac{Y_e\rho}{\rho_0}\right)^{2/3}T_9\:\mathrm{erg\,cm^{-3}\,K^{-1}}.
\end{eqnarray}
where $Y_e$ is the electron fraction.

Recently, Rajagopal and Wilczek ~\citep{r2001} have shown that the quark matter in the color-flavor locked phase of QCD, which occurs for large gaps $(\Delta \sim 100MeV$), is rigorously electrically neutral, despite the unequal quark masses, and even in the presence of electron chemical potential. In other words, $Y_e$ = 0, and hence $c_e$ = 0. However, in order to have this electrical neutral situation the following condition must be satisfied
\begin{equation}
\mu \, > \,\frac{3m_s^2}{2\sqrt{2}\Delta}.
\label{eq:neutral}
\end{equation}
where $\mu $ is the baryon chemical potential and $\mu _s$ is the strange quark mass.
Realistically, a strange star must have a density profile. If the above condition is satisfied in the core, it may not be satisfied near the surface, where the density is equal to 4B. Taking the typical values of baryon density near the strange star surface $n_b \approx 0.3$fm$^{-3}$, the baryon chemical potential is $\sim 280$MeV and the right hand side of the equation gives $\sim 520MeV$ with $m_s$ = 220 MeV. Therefore we believe that $Y_e$ will not be zero throughout a strange star even if its core is in color-flavor locked phase.
For simplicity, we treat $Y_e$ as a parameter. 

Neutrino is emitted from quark URCA process:
\begin{eqnarray}
d&\rightarrow&u+e+\bar\nu_e	\label{eq:dbeta1}\\
u+e&\rightarrow&d+\nu_e.	\label{eq:dbeta2}
\end{eqnarray}
Using the assumption of eq.~(\ref{eq:pf}), \citet{iw80} obtained the
neutrino
emissivity:
\begin{equation}
\epsilon _d\simeq 8.8\times 10^{26}\alpha_c\left( \rho/\rho_0\right) Y_e^{1/3} T_9^6\,\mathrm{erg\,cm^{-3}\,s^{-1}},
\label{eq:ev}
\end{equation}
where $\alpha_c$ is the strong coupling constant.
In superfluid state, i.e. $T\leq T_c$, it is suppressed by a factor of
$\exp(-\Delta/T)$. Note that eq.~(\ref{eq:ev}) is an approximation to the
exact form \citep{du83}:\\ \begin{widetext}
\begin{equation}
\epsilon_d=\frac{457\pi}{840\hbar^{10} c^9}G_F^2\cos^2\theta_C\left( 1-\frac{1}{a}\cos\theta_{ue}\right )
\frac{1}{a^2}P_{Fu}\,P_{Fd}\,P_{Fe}\,(k_BT)^6\,\mathrm{erg\,cm^{-3}\,s^{-1}},
\label{eq:ev2}
\end{equation}\\
\end{widetext}
where $G_F$ is the Fermi weak coupling constant, $\theta_C$ is the Cabibbo angle,  $k_B$ is the Boltzmann constant, $a=(1-\frac{2\alpha_c}{\pi})^{-1/3}$, $P_{Fu}$, $P_{Fd}$, $P_{Fe}$ are Fermi momenta of $u$, $d$ quarks and electrons respectively, $\theta_{ue}$ is the angle between the $u$ and $e$ momenta, which satisfies:
\begin{equation}
\cos\theta_{ue}=\frac{P^2_{Fd}-P^2_{Fu}-P^2_{Fe}}{2P_{Fu}P_{Fe}}.
\end{equation}
The condition for reactions~(\ref{eq:dbeta1}) and~(\ref{eq:dbeta2}) to occur is $\left|\, \cos\theta_{ue}\right | \leq1$. We solved the equations numerically and found that in the density range we are interested, the reactions can always occur.

On the other hand, we also considered the similar reactions for $s$ quarks,
\begin{eqnarray}
s&\rightarrow& u + e +\bar\nu_e \label{eq:sbeta1}\\
u + e&\rightarrow& s+ \nu_e.
\label{eq:sbeta2}
\end{eqnarray}
Numerical results showed that the value of $|\cos\theta_{ue}'|$, which is defined by
\begin{equation}
\cos\theta_{ue}'=\frac{P^2_{Fs}-P^2_{Fu}-P^2_{Fe}}{2P_{Fu}P_{Fe}}
\end{equation}
is smaller than 1 only if the density is as high as $4\times10^{16}\,\mathrm{g\,cm^{-3}}$. Therefore, reactions~(\ref{eq:sbeta1}) and~(\ref{eq:sbeta2}) do not exist in the star. Here, we have also ignored the neutrino emissivity due to the photon-gluon mixing, which can generate the a massive photon-gluon excitation ~\citep{b2000}.  It is because this excitation is only important for the temperature higher than 70 MeV and our cooling calculations of quark stars start with temperature about 10MeV.

The most important feature of Cloudy Bag strange stars is the emission of pions. Due to the collision between quarks and the bag, pions are created at the stellar surface, which represents a conversion of the quark kinetic energy to the mass of the pion cloud. Some of the more energetic pions may even escape from the star carrying away substantial energy. The pions are assumed to be decayed once left the stellar surface for a distance larger than $1/\mu$. The decay channel is
\begin{eqnarray}
\pi^0 \rightarrow 2\gamma &\leftrightarrow& \;e^+e^- \\
\pi^\pm  \rightarrow  \mu^\pm & +&\nu_\mu \nonumber \\
\mu^\pm & \rightarrow & e^\pm + \nu_e + \nu_\mu.
\end{eqnarray}
As the decay rate is nearly 100\%, it is an efficient cooling mechanism and the final product is a fireball of $e^+e^-$ pairs and photons.

We estimated the escape velocity for pions to leave the region of $1/\mu$ from stellar surface. Numerical results showed that this value is very small such that almost every pion can escape and decay. Hence, we can assume all pions produced will decay. The pion emissivity is given by:
\begin{equation}
L_\pi = \rho _\pi\:v_\pi\:4\pi R^2,
\end{equation}
where $\rho_\pi$ is the energy density of pion field at stellar surface, which is fixed by the axial current conservation in Cloudy Bag Model. Speed of pions emitted, $v_\pi$, is estimated by $\sqrt{\frac{2k_BT}{\mu}}$, where $\mu$ is the pion mass. Therefore,
\begin{equation}
L_\pi =  \rho_\pi\:\sqrt{\frac{2k_BT}{\mu}}\:4\pi R^2.
\end{equation}
Put in $\rho_\pi=7.1\times 10^{31}\,\mathrm{erg\,cm^{-3}}$, $R=10^6\,$cm, we get:
\begin{equation}
L_\pi=9.5\times10^{53}\,T_9^{1/2}\,\mathrm{erg\,s^{-1}}.
\end{equation}
In superfluid state, since the collision between quarks and the bag is suppressed. By a similar argument, pion emissivity is also reduced by the factor of $\exp(-\Delta/T)$.

In eq.(34), we show that pions can carry away as much as $\sim 10^{54}$ erg/s. However, the boundary condition requires the density of pion is only $10^{11} g/cm^3$, from which we have estimated that the entire strange star will be covered by a pion cloudy with mass $\sim  10^{-22} M_{\odot}$. The key reasons why pions with such low masses can carry away that much energy are: (1)The thermal conductivity of compact object like a strange star is extremely high, therefore it is a good approximation to assume that the entire strange star is isothermal. In fact the characteristic time scale for the thermal equilibrium is $R/v_s$ where $R$ is the radius of the star and $v_s$ is the speed of sound, which is $c/\sqrt{3}$. This time scale is $5.7\times10^{-5}s$. Therefore the thermal energy can quickly transport from core to the surface. (2)In order to carry away $\sim 10^{54}$erg/s, the pion production rate $\dot{N}_{\pi}$ must be larger than $\dot{N}_{\pi}> \frac{10^{54} erg/s }{m_{\pi}c^2}\sim 10^{58}/s$. Since pions are produced when quarks collide the boundary of the bag, we can estimate that the collision rate between quarks and the boundary is $n_q v_q 4\pi R^2 \sim 10^{62}/s$ , where $n_q$ is the quark number density and  $v_q$ is the quark velocity. We have assumed that the quark mass density is 4B and quarks are moving relativistically. We can see that most pions will be re-absorbed by the boundary in order to maintain the pion density at the boundary. Only a small fraction of pions can escape.

The last cooling process we considered in Cloudy Bag strange stars is blackbody radiation. It is given by the well-known formula:
\begin{equation}
L_{bb}=4\pi R^2\sigma T^4,
\end{equation}
where $\sigma$ is the Stefan-Boltzmann constant. 
However, it has been pointed out that bare strange stars are very poor radiators of thermal photons with energy less than 20MeV. This is because the plasma frequency ($\omega_p$) that is related to particle density of quarks is very high even at the surface of strange stars (Alcock et al. 1986). On the other hand, Chmaj et al. (1991) argue that low-energy photons with energy less than the plasma frequency can still leave the strange star surface due to the non-equilibrium quark-quark bremstralung radiation in the surface layer with the thickness of $\sim c/\omega_p \approx 10$fm. Nevertheless the efficiency of this non-equilibrium blackbody radiation is only $\sim 10^{-4}$ of $L_{bb}$. 
As it is a bare strange star in our case, the surface temperature is high and the energy of the blackbody radiation is large. Besides, in superfluid state, other cooling processes are suppressed, so this non-equilibrium blackbody radiation will be a very important cooling mechanism at the later stage.

Here, we want to remark that Eq.34 and Eq.35 represent two completely different physical processes. The production of pion is the requirement of chiral symmetry, which is believed to be a good symmetry for strong interaction, and they are produced outside the star due to the collisions between quarks and the bag boundary. On the other hand, the reduced blackbody radiation (Chmaj et al. 1991) results from the collisions among quarks. They are electromagnetic waves produced very near the stellar surface but yet inside the star and these EM waves leak out the star through the bag boundary. 

In order to have a more realistic calculation,
the effects of thermal equilibrium radiation and emission of $e^+e^-$
pairs, as suggested by~\citet{us01}, were also discussed. In this theory,
hot
strange matter is filled with electromagnetic waves in thermodynamic
equilibrium with quarks. The energy flux of thermal equilibrium photons
radiated from a bare strange star surface is given by~\citet{ch91}:
\begin{equation}
F_{eq}=\frac{\hbar}{2}\int^\infty_{\omega_p}d\omega \frac{\omega(\omega^2-\omega^2_p)g(\omega)}{\exp(\hbar\omega/k_BT_s)-1},
\end{equation}
where $\hbar\omega_p\simeq 20-25$MeV, $T_s$ is the surface temperature,
\begin{equation}
g(\omega)=\frac{1}{2\pi^2}\int^{\pi/2}_0d\theta\sin\theta\cos\theta D(\omega,\theta)
\end{equation}
and $D=1-R(R_\bot+R_\|)/2$ is the coefficient of radiation transmission from strange matter to vacuum.
\begin{equation}
R_\bot=\frac{\sin^2(\theta-\theta_0)}{\sin^2(\theta+\theta_0)} \qquad \qquad R_\|=\frac{\tan^2(\theta-\theta_0)}{\tan^2(\theta+\theta_0)}
\end{equation}
\begin{equation}
\theta_0=\arcsin[\sin\theta\sqrt{1-(\omega_p/ \omega)^2}].
\end{equation}

It was pointed out by~\citet{us98} that the strong electric field at the
surface of a
strange star may be a powerful source of $e^+e^-$ pairs. The production
rate is~\citep{us01}
\begin{equation}
\dot n_\pm\simeq10^{39}\left(\frac{T_s}{10^9\mathrm{K}}\right)^3\exp\left[-11.9\left(\frac{T_s}{10^9\mathrm{K}}\right)^{-1}\right]J(\xi) \,\mathrm{s^{-1}},
\end{equation}
where
\begin{equation}
J(\xi)=\frac{1}{3}\frac{\xi^3\ln(1+2\xi^{-1}}{(1+0.074\xi)^3}+\frac{\pi^5}{6}\frac{\xi^4}{(13.9+\xi)^4}
\end{equation}
\begin{equation}
\xi=2\sqrt{\frac{\alpha}{\pi}}\frac{\varepsilon_F}{k_BT_s}\simeq0.1\frac{\varepsilon_F}{k_BT_s},
\end{equation}
where $\alpha=1/137$ is the fine-structure constant and $\varepsilon_F\simeq$18\,MeV.
The energy flux of $e^+e^-$ pairs from unit surface is given by
\begin{equation}
F_\pm\simeq\varepsilon_\pm\dot n_\pm,
\end{equation}
where $\varepsilon_\pm\simeq m_ec^2+k_BT_s$ is the mean energy of created particles.

The cooling mechanisms of MIT strange stars are the same as those of Cloudy Bag Strange stars except without the pion emission. As for comparison, the cooling of neutron stars is also discussed.
Following the work by~\citet{ch92}, we considered a simple case which only
has
neutrino emission and blackbody radiation for cooling. $T_c$ is taken to
be $3.2\times 10^9$\,K for the transition to superfluid
state~\citep{ta71}.
The detail relation between the core temperature $T$ and the surface temperature $T_s$ can be obtained by solving the energy transport equations numerically (e.g. Tsuruta 1979) and is found to be approximately related by a simple power law  $T_s = (10T)^{\frac{2}{3}}$(see Shapiro and Teukolsky 1983, p.330). Obviously, this simple relation cannot take into account the gravitational effect. \citet{gu83} have carried out a more detail analysis and found that:
\begin{equation}
T_8=1.288(T^4_{s6}/g_{s14})^{0.455},
\label{eq:tstc}
\end{equation}
where $g_{s14}$ is the surface gravity in the unit of $10^{14}\, \mathrm{cm\,s^{-1}}$.

Specific heat of neutron matter is given by~\citep{ma79}:
\begin{equation}
c_n=2.3\times10^{39}M_*\rho_{14}^{-2/3}T_9 \,\mathrm{erg \,K^{-1} \qquad for}\ T>T_c,
\end{equation}
where $M_*$ is the stellar mass in the unit of solar mass, $\rho_{14}=\rho/10^{14}\,\mathrm{g\, cm^{-3}}$. In superfluid state, it is modified to be:\\
\begin{widetext}
\begin{equation}
c_n^{\mathrm{sf}}= 3.15c_n\left( \frac{T_c}{T}\right) e^{-1.76T_c/T}\left[\,2.5 -1.66\left(\frac{T}{T_c}\right) +3.64\left(\frac{T}{T_c}\right)^2\right] \ \mathrm{for} \ T\leq T_c,
\end{equation}\\
\end{widetext}
which is exactly the same formula as eq.~(\ref{eq:csf}). Specific heat of electrons is
\begin{equation}
c_e=1.9\times 10^{37}M_*\rho_14^{1/3}T_9 \, \mathrm{erg \, K^{-1}}.
\end{equation}

Neutrino luminosity consists of two parts: the first part is

electron-proton Coulomb scattering $(e+p\rightarrow e+p+\nu+\bar\nu)$ in
the crust~\citep{fe69}:
\begin{equation}
L^{\mathrm{cr}}_\nu=1.7\times 10^{39}M_*(M_\mathrm{cr}/M)T^6_9 \, \mathrm{erg\,s^{-1}},
\end{equation}
where $(M_\mathrm{cr}/M)$ is the fractional mass of the crust $\approx5\%$.

\newpage \noindent The second part is neutrino bremsstrahlung process $(n+n\rightarrow n+n+\nu+\bar\nu)$, which is given by:
\begin{widetext}
\begin{eqnarray}
L^{\mathrm{nn}}_\nu&=&4.3\times 10^{38}\rho_{14}^{1/3}\,T^8_9 \, \mathrm{erg \, s^{-1} \qquad \qquad for} \ T>T_c \\
&=&0 \hspace{5.75cm} \mathrm{for} \ T\leq T_c.
\end{eqnarray}
\end{widetext}

Another case of neutron star is similar, but with the addition of a rapid cooling mechanism. Direct URCA process
\begin{eqnarray}
n&\rightarrow&p+e^-+\bar\nu_e \\
p+e^-&\rightarrow&n+\nu_e
\end{eqnarray}
is known to be the most efficient cooling way in neutron stars. At high
temperature, the neutrino emissivity can be higher than that from pion
condensation and quark URCA process by one to two orders of magnitude.
\citet{la91} suggested that this process may occur in neutron stars if the
proton
concentration exceeds a certain value. They estimated the neutrino emissivity to be
\begin{equation}
\epsilon_{URCA}=4.00\times10^{27}(Y_en/n_0)^{1/3}T^6_9 \, \mathrm{erg\,cm^{-3}\,s^{-1}},
\end{equation}
where $Y_e$ is the electron fraction, it is taken to be 0.1 in our calculation and $n_0=2.8\times 10^{28}\,\mathrm{g\,cm^{-3}}$ is the nuclear saturation density. We also include the pion condensation (see Shapiro and Teukolsky 1983 P.323) as anther possible rapid cooling mechanism for neutron stars. 

\subsection{Cooling Curves}
The cooling behaviors and the surface luminosities of 1.4$M_\odot$ strange stars and neutron stars are shown from fig.8 to fig.14. The Cloudy Bag strange star is the one obtained using best-fit parameters, i.e. $B^{1/4}=120$\,MeV and $m_s=220$\,MeV; the MIT one is using the linear EOS and the neutron star is the one with typical mass of 1.4$M_\odot$ and radius of 10\,km. Strange stars are bare while the neutron star has a crust of mass about 5\% of the total stellar mass.

The initial temperature was chosen to be $10^{11}$\,K, which is a typical
temperature when the star is born. Due to gravitational redshift, the
surface temperature and luminosity observed at infinity ($T^\infty_s$ and
$L^\infty$) are related to their value at stellar surface ($T_s$ and $L$)
by~\citet{ts98}:
\begin{eqnarray}
T^\infty_s=e^{\phi_s/c^2}T_s \\
L^\infty=e^{2\phi_s/c^2}L,
\end{eqnarray}
where the surface redshift factor is
\begin{equation}
e^{\phi_s/c^2}=\sqrt{1-\frac{2GM}{Rc^2}}.
\end{equation}
$M$ is the gravitational mass of the star, $R$ is the stellar radius and $G$ is the gravitational constant.

\begin{figure}[h]
    \centerline{\epsfxsize=12cm
                \epsfbox{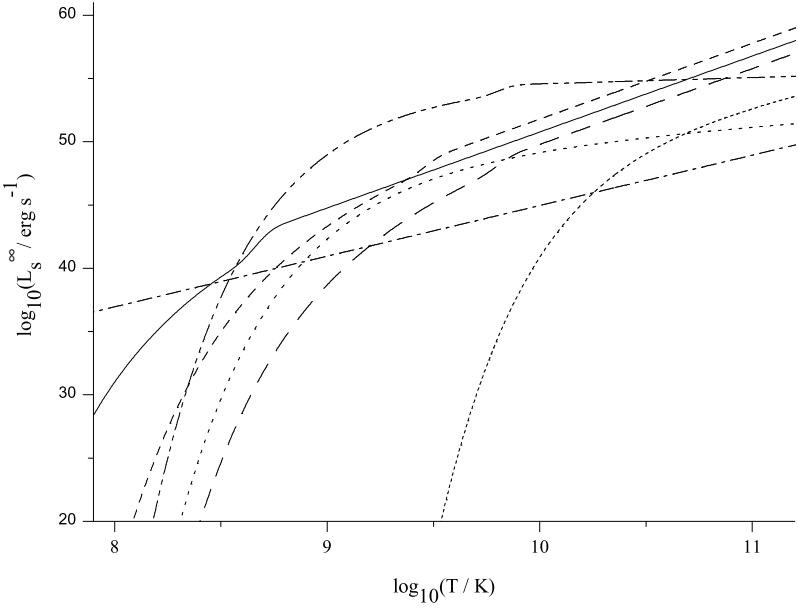}}
	\caption{Luminosities of various emission mechanisms observed at infinity as a function of temperature. The dash-dot-dotted line is the pion luminosity; the dashed line is the neutrino luminosity from quark URCA process; the dotted line is the $e^+e^-$ pair emission luminosity; the short dotted line is the equilibrium radiation and the dash-dotted line is the non-equilibrium blackbody radiation. The short dashed line and solid line are neutrino luminosities for direct URCA process and pion condensation respectively for neutron stars.}
	\label{fig:8}
\end{figure}

Figure~\ref{fig:8} is the luminosities of various emission mechanisms observed at infinity as a function of temperature. The calculations are based on Cloudy Bag strange star with $\Delta=1$\,MeV, $Y_e=10^{-3}$. As a comparison, two cooling mechanisms for neutron stars are also shown. We considered the direct URCA process with $Y_e=10^{-3}$ and $T_c=3.2\times 10^9$\,K. For pion condensation model, $Y_e=10^{-3}$ and $T_c=5\times 10^8$\,K.

According to the graph, energy released in direct URCA process (short dashed line) is highest at high temperature. Then the luminosity by pion condensation (solid line) follows. quark URCA emission follows. When temperature drops below $3\times10^{10}$\,K, pion emission (dash-dot-dotted line) becomes the most efficient cooling mechanism and dominates over a wide range of temperature. When temperature drops further, the star is in superfluid state. Most of the emissions are suppressed and only non-equilibrium blackbody radiation dominates. Although thermal equilibrium radiation and $e^+e^-$ pair emission remain unaffected, their values are small and negligible in low temperature.

\begin{figure}[h]
    \centerline{\epsfxsize=12cm
                \epsfbox{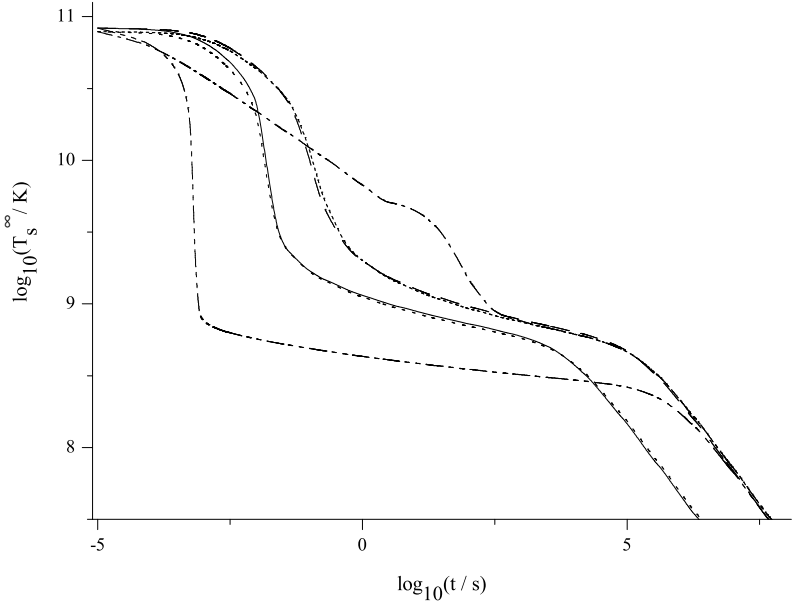}}
		\caption{Surface temperature of different strange stars obvserved at infinity. For MIT strange stars, the dash-dotted line is $\Delta=1$\,MeV and $Y_e=10^{-3}$; the short dotted line is $\Delta=100$\,MeV and $Y_e=10^{-3}$; the dotted line is $\Delta=100$\,MeV and $Y_e=10^{-5}$. For Cloudy Bag strange stars, the dash-dot-dotted line is $\Delta=1$\,MeV and $Y_e=10^{-3}$; the dashed line is $\Delta=100$\,MeV and $Y_e=10^{-3}$; the solid line is $\Delta=100$\,MeV and $Y_e=10^{-5}$.}
	\label{fig:9}
\end{figure}

The surface temperature of different kinds of stange stars observed at infinity is plotted in figure~\ref{fig:9}. We first considered a Cloudy Bag strange star with $\Delta=1$\,MeV and $Y_e=10^{-3}$ (the dash-dot-dotted line). As pion emission is an efficient cooling mechanism, temperature of the Cloudy Bag strange star drops rapidly at the beginning. From the figure, temperature drops from $10^{11}$\,K to $10^{9}$\,K within 1\,ms. A rough estimation of energy released:
\begin{equation}
\Delta E\sim N_Bk\Delta T\sim 10^{52}\, \mathrm{erg},
\end{equation}
where $N_B\sim10^{57}$ is the total baryon number of the star. 

It has been suggested that the conversion of neutron stars to strange
stars could be the origin of $\gamma$-ray bursts~\cite{ch96} and 40\% of
the deconfinment energy ($\sim2\times10^{52}$\,erg) will be converted into
$e^+e^-$ pairs and photons fireball via neutrino-antineutrino
annihilation. However, in our model, the evaporation of pions is a more
efficient heat dissipation mechanism. The pion decay products form a
fireball of $e^+e^-$ and photons directly instead of via
neutrino-antineutrino annihilation. Therefore the fireball can carry away
100\% of the deconfinment energy. Since both the time scale and order of
magnitude of energy released is similar to typical $\gamma$-ray
bursts~\cite{ch96}, we suggested that the rapid cooling of Cloudy Bag

strange stars may be a possible energy source of $\gamma$-ray bursts.

When the core temperature drops below $10^9$\,K, the cooling slows down. In this stage, the strange matter becomes superfluid. Therefore, emissivity of neutrino and pion are reduced by a factor of $\exp(-\Delta/T)$. Finally, the star is cooled by blackbody radiation only.

For a MIT strange star with $\Delta$=1\,MeV and $Y_e=10^{-3}$ (the dash-dotted
 line), it is cooled by neutrino emission, thermal equilibrium radiation, $e^+e^-$ emission and blackbody radiation. Thus, the cooling rate is much slower than that of a Cloudy Bag strange star, which emitts neutrinos as well as pions.

Similarly, while temperature is lower than $10^9$\,K, all cooling mechanisms become negligible except blackbody radiation. This is why the long term thermal evolution is the same in both MIT and Cloudy Bag strange stars.

In the case of $\Delta$=100\,MeV, temperature of the stars can never exceed the critical temperature $T_c$. In other words, they are already in superfluid state since born. Hence pion and neutrino emissivity are very small. The cooling behaviour of Cloudy Bag and MIT strange stars are similar, since they are mainly cooled by the smae mechanisms.

Note that for $\Delta$=100\,MeV, stars with $Y_e=10^{-5}$ cool faster than those with $Y_e=10^{-3}$. It is because in superfluid state, specific heat mainly comes from the contribution of electrons. Therefore, if $Y_e$ is smaller, the star can cool faster. 

It is interesting to see that for MIT strange stars, although the one with $\Delta$=100\,MeV cools slower, its surface temperature is lower than that of $\Delta$=1\,MeV after 1s. This can be explained by the specific heat, since it depends on $\Delta$. Under the same temperature, the specific heat is smaller for a larger gap. Hence, the temperature drops faster.

For the long-term behavior, since it is dominated by the non-equilibrium blackbody radiation. They only depend on $Y_e$, which is related to the specific heat.

\begin{figure}[h]
    \centerline{\epsfxsize=12cm
                \epsfbox{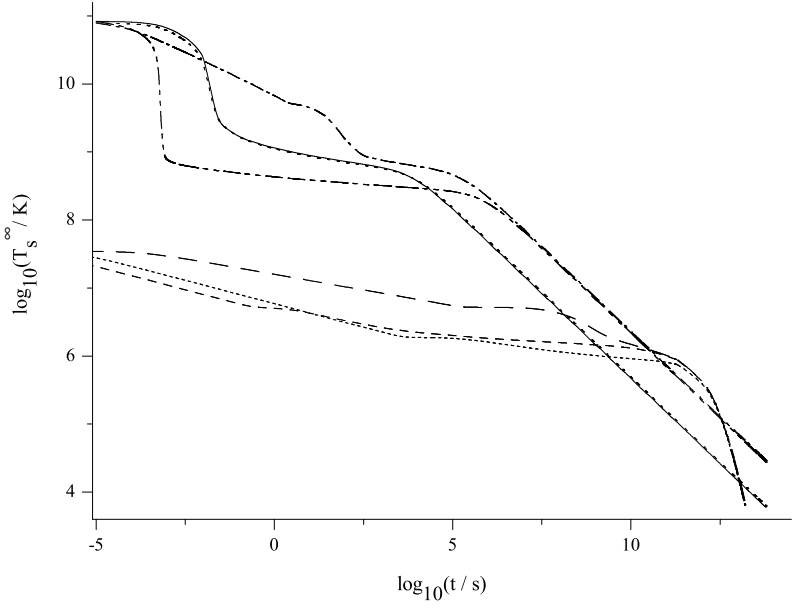}}
		\caption{Surface temperature of strange stars and neutron stars obvserved at infinity. For MIT strange stars, the dash-dotted line is $\Delta=1$\,MeV and $Y_e=10^{-3}$; the dotted line is $\Delta=100$\,MeV and $Y_e=10^{-5}$. For Cloudy Bag strange stars, the dash-dot-dotted line is $\Delta=1$\,MeV and $Y_e=10^{-3}$; the solid line is $\Delta=100$\,MeV and $Y_e=10^{-5}$. For neutron stars, the dashed line is for ordinary one, the short dashed line is for the one with direct URCA process and the short dotted line is for the one with pion condensation.}
	\label{fig:10}
\end{figure}

Figure~\ref{fig:10} shows the surface temperature of strange stars and neutron stars observed at infinity. Since neutron stars are covered by crusts, which act as thermal blankets to reduce the energy lost rate. At the beginning, although the core temperature of neutron stars is higher, their surface temperature is lower than strange stars by a lot. This is due to eq.~(\ref{eq:tstc}), surface temperature is different from core temperature by several orders of magnitude. Since neutron stars cool slowly, the surface temperature is eventually higher than that of strange stars at $\sim 10^{8}$\,s for $\Delta$ = 100MeV and at $\sim 10^{11}$\,s for $\Delta$ = 1MeV respectively.

\begin{figure}[h]
    \centerline{\epsfxsize=12cm
                \epsfbox{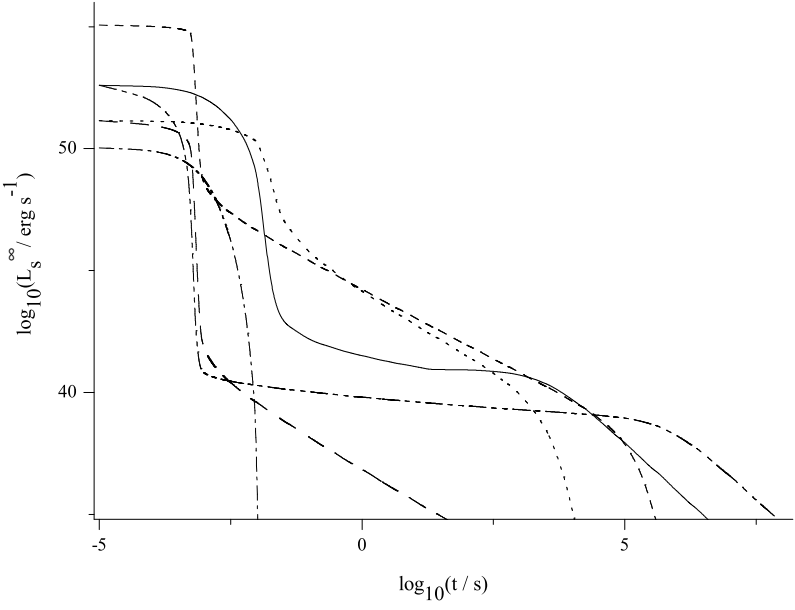}}
	\caption{Luminosities of electromagnetic radiation from various emission mechanisms for Cloudy Bag strange stars observed at infinity. For $\Delta=1$\,MeV and $Y_e=10^{-3}$, the short dashed line is the pion luminosity; the dash-dot-dotted line is the thermal photon luminosity; the dashed line is the $e^+e^-$ pair production, which will annihilate to MeV photons. For $\Delta=100$\,MeV and $Y_e=10^{-5}$, the dash-dotted line is the pion luminosity, which will decay to 100 MeV photons; the solid line is the thermal photon luminosity; the dotted line is the $e^+e^-$ pair production.}
	\label{fig:11}
\end{figure}

\begin{figure}[h]
    \centerline{\epsfxsize=12cm
                \epsfbox{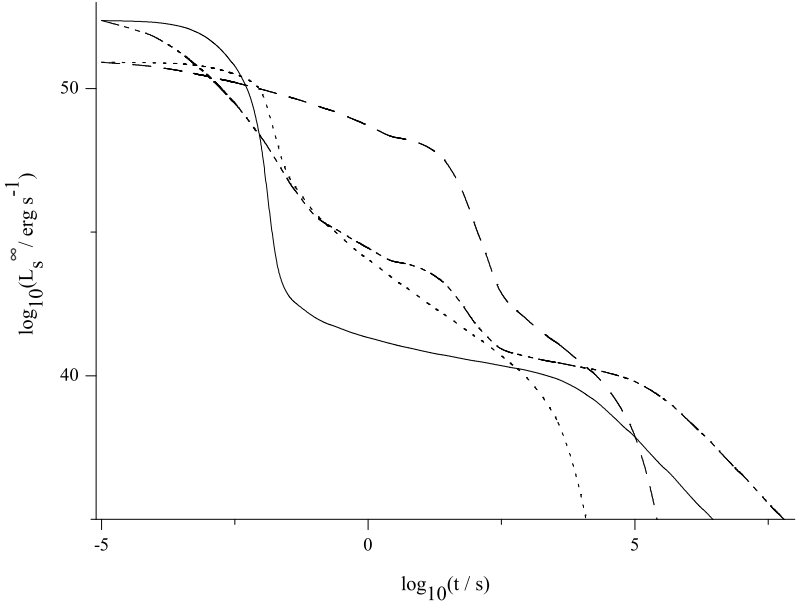}}
	\caption{Luminosities of electromagnetic radiation from various emission mechanisms for MIT strange stars observed at infinity. For $\Delta=1$\,MeV and $Y_e=10^{-3}$, the dash-dot-dotted line is the thermal photon luminosity; the dashed line is the $e^+e^-$ pair production. For $\Delta=100$\,MeV and $Y_e=10^{-5}$, the solid line is the thermal photon luminosity; the dotted line is the $e^+e^-$ pair production.}
	\label{fig:12}
\end{figure}

To investigate the observational properties, the luminosities of electromagnetic radiation from various emission mechanisms for Cloudy Bag strange stars and MIT strange stars were plotted in figure~\ref{fig:11} and~\ref{fig:12} respectively. Photon luminosity plotted in the graphs consists of thermal equilibrium photons and non-equilibrium blackbody radiation. Note that in both graphs, photon luminosity and $e^+e^-$ pair production start at the same point. Since they are independent of superfluidity, they are the same for all kinds strange stars. They become different in later time because of different evolution of the star temperature. 

For Cloudy Bag strange stars with $\Delta=1$\,MeV and $Y_e=10^{-3}$ (fig.~\ref{fig:11}), pion emission (short dashed line) dominates over a wide range of time until $10^5$\,s. Hence electromagnetic radiation mainly comes from the decay of pion in this case.

In the case of Cloudy Bag strange stars with $\Delta=100$\,MeV and $Y_e=10^{-5}$, pion emission (dash-dotted line) is suppressed by superfluidity and becomes negligible. Therefore electromagnetic radiation mainly comes from blackbody radiation, equilibrium radiation and annihilation of $e^+e^-$ pairs.

For MIT strange stars with $\Delta=1$\,MeV and $Y_e=10^{-3}$ (fig.~\ref{fig:12}), electromagnetic radiation mainly comes from $e^+e^-$ pairs (dashed line) from $10^{-2}$\,s to $10^5$\,s, while photon emission (dash-dot-dotted line) dominates after $10^5$\,s. Similar for the case of $\Delta=100$\,MeV and $Y_e=10^{-5}$, but the photon luminosity (solid line) dominates after $10^3$\,s.

\begin{figure}[h]
    \centerline{\epsfxsize=12cm
                \epsfbox{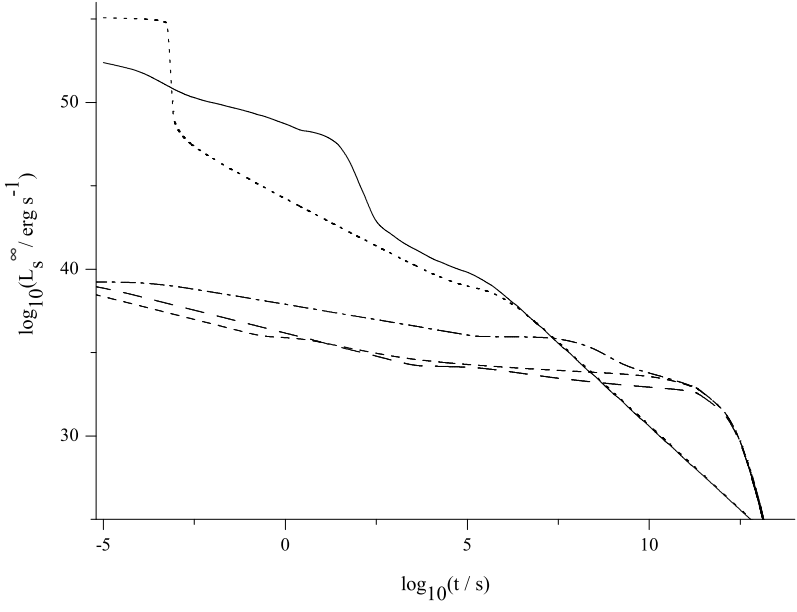}}
	\caption{Total electromagnetic radiation (including thermal photons, MeV photons from the annihilation of $e^+e^-$ pairs and 100MeV photons from the decay of pions) from strange stars with $\Delta=1$\,MeV, $Y_e=10^{-3}$ and neutron stars observed at infinity. The solid line is MIT strange star; the dotted line is Cloudy Bag strange star; the dash-dotted line is ordinary neutron star; the short dashed line is neutron star with direct URCA process; the dashed line is neutron star with pion condensation.}
	\label{fig:13}
\end{figure}

\begin{figure}[h]
    \centerline{\epsfxsize=12cm
                \epsfbox{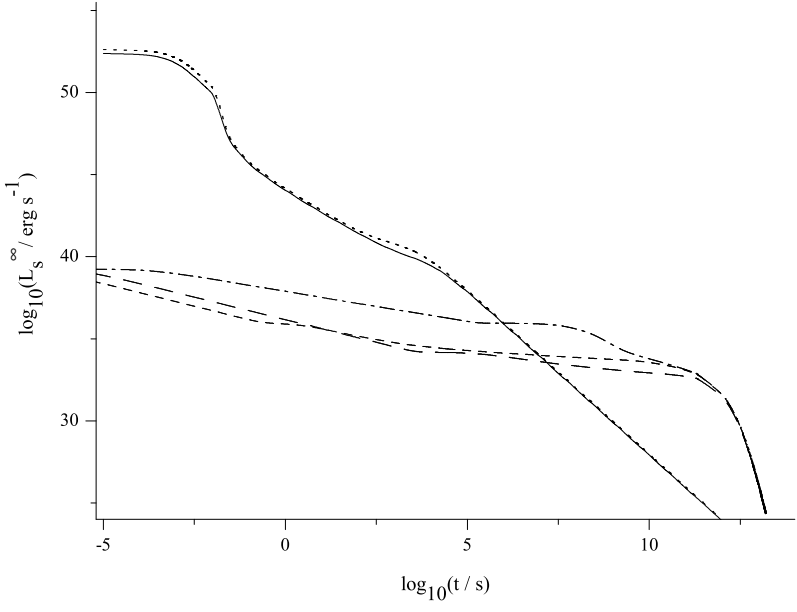}}
	\caption{Total electromagnetic radiation from strange stars with $\Delta=100$\,MeV, $Y_e=10^{-5}$ and neutron stars observed at infinity. The solid line is MIT strange star; the dotted line is Cloudy Bag strange star; the dash-dotted line is ordinary neutron star; the short dashed line is neutron star with direct URCA process; the dashed line is neutron star with pion condensation.}
	\label{fig:14}
\end{figure}

The total electromagnetic radiation from different strange stars and neutron stars are plotted in figure~\ref{fig:13} and~\ref{fig:14}. Strange stars in figure~\ref{fig:13} were calculated using $\Delta=1$\,MeV and $Y_e=10^{-3}$. From the figure, the Cloudy Bag strange star (dotted line) is most luminous at the beginning. Then after 1ms, it transits into superfluid state and the pion luminosity drops drastically. At this time, the luminosity of MIT strange star with $\Delta=1$\,MeV, $Y_e=10^{-3}$ (solid line) is highest. At $10^2$\,s, it also drops due to superfluidity. Around $10^8$\,s, the luminosity of strange stars drop below neutron stars.

For strange stars with $\Delta=100$\,MeV and $Y_e=10^{-5}$ (figure~\ref{fig:14}), both MIT (solid line) and Cloudy Bag (dotted line) strange stars behave similarly, as pion emission is highly suppressed. In this case, the luminosity drops below that of neutron stars at $10^6$\,s. As compared to previous graph, $Y_e$ is smaller for strange stars, therefore specific heat is smaller and the stars cools faster. Hence the luminosity drops below that of neutron stars at an earlier time.


\section{Discussion}
We have used the Cloudy Bag model to describe the properties of strange stars.
The model parameters, i.e., the bag constant $(B=(120\, {\rm MeV})^4)$ 
and the strange quark mass $(m_s = 220 {\rm\, MeV})$, are fixed by comparing with hadron masses.
We found that the star is surrounded by a thin pion cloud, which is not significant in the stellar mass and radius. Hence, stellar properties are similar in the Cloudy Bag and MIT case, except that the former has slightly larger radius and smaller mass due to a larger $B$. The maximum mass of Cloudy Bag strange stars is about $1.9 \,M_\odot$.

For the rotation properties of Cloudy Bag strange stars, the approximate formulae were employed. We found that the Kepler limit for Cloudy Bag strange stars with the best-fit parameters is about 6500\,s$^{-1}$, which is lower than the MIT one by $\sim 30\%$. The maximum mass of a stable rotating star is about $2.3\,M_\odot$, similar to the value $2.4\,M_\odot$ of a MIT strange star.

The cooling properties of bare strange stars sensitively depend on the gap energy. For small gap energy ($\Delta$ = 1MeV), the cooling properties of MIT strange stars and Cloudy strange stars are very much different because the pion emission can cool the strange star rapidly. For the large gap energy ($\Delta$ = 100MeV), the quarks become superconducting immediately after the star is formed, and the pion emission is totally suppressed. MIT strange stars and Cloudy strange stars become no difference. Furthermore, large gap energy also creates the effect of color-flavor locked, which significantly reduces the electron concentration and its capacity. This makes the cooling curves lower than those with small gap energy. From fig13 and fig14, we can see that
the surface luminosity of a bare strange star is higher than that of a neutron star until $10^6s$ and $10^8s$ for ($\Delta$ = 100MeV) and ($\Delta$ = 1MeV) respectively. After this period, the surface luminosity of a bare strange star becomes lower than that of a neutron star even rapidly cooling mechanisms, e.g. direct URCA process or pion condensation, exist in the core of neutron stars. Hence, the cooling behavior may provide a possible way to distinguish a compact object between a neutron star, MIT strange star and Cloudy Bag strange star in observations.

It is interesting to note that although the MIT and Cloudy strange stars are hotter than neutron stars before  $\sim 10^{8}$\,s for $\Delta$ = 100MeV and  $\sim 10^{11}$\,s for $\Delta$ = 1MeV respectively(cf. fig10), the thermal radiation of neutron stars are still stronger than those of strange stars 
Until until $10^6s$ and $10^8s$ for ($\Delta$ = 100MeV) and ($\Delta$ = 1MeV) respectively. It is because the cooling mechanism at the later stage is dominated by the black body radiation, which is strongly suppressed for strange stars as pointed in section IVB.

>From fig13 and fig14, we can see that the cooling curves of a MIT strange star and a cloudy strange star are the same when the cooling mechanism is dominated by blackbody radiation. However, the radiation properties of a MIT strange star and a cloudy strange star are very much different in the early stage ( t $< 10^6s$ ) for small gap energy case. The cloudy strange star is first cooling very rapidly by emitting pions, which decay to 100MeV photons and its temperature drops quickly below that of MIT strange star. Therefore the MIT strange star is cooling mainly by emitting electrons/positrons. The total energy of a cloudy strange star carried away by 100MeV photons is $\sim 10^{52}ergs$, which makes the evaporation of pions a very efficient $\gamma$-ray burst mechanism. 

Finally, we want to remark that the cooling curves of strange stars are lower than those cooling curves of neutron stars, which are consistent with the current observed data ~\citep{ts98}. Therefore we actually suggest that the cooling curves may be one possible way to differentiate strange stars from neutron stars. For example, strange stars can behave as pulsars if they have strong magnetic field~\citep{xu2001}~\citep{xb2001}. Therefore their ages can be estimated by their period and period derivative. If these pulsars are found to have thermal emission lower than that of the standard cooling neutron stars, then we can further to see if there are any e$^{\pm}$ annihilation lines, which are red shifted larger than those of neutron stars (cf. fig. 5). The features of thermal X-ray emitted from these pulsars associated with bare strange stars are also very different from those of neutron stars~\citep{xu2002}. 

This work is partially supported by a RGC grant of the Hong Kong 
Government.

\begin{table*}[h]
\begin{center}
\begin{tabular}{c|ccccccc}
$B^{1/4}\,\mathrm{(MeV)}$ &&100 & 110 & 120 & 130 & 140 & 150 \\ \hline 
\mbox{r.m.s. error (MeV)} &&208.6 & 149.9 & 143.8 & 195.6 & 274.6 & 363.8
\end{tabular}
\caption{The average r.m.s. error in the fitting for baryons which do not contain strange quarks.}	\label{table:b}
\end{center}
\end{table*}

\begin{table*}[h]
\begin{center}
\begin{tabular}{c|ccccccc}
$m_s\,\mathrm{(MeV)}$ &&200 & 220 & 240 & 260 & 280 & 300 \\ \hline
\mbox{r.m.s. error (MeV)} && 114.0 & 111.9 & 114.1 & 120.8 & 131.4 & 145.4
\end{tabular}
\caption{The average r.m.s. error in the fitting for baryons which contain one or more strange quarks.}	\label{table:ms}
\end{center}
\end{table*}

\begin{table*}[h]
\begin{center}
\begin{tabular}{cc|cc|ccc}
&\multicolumn{3}{r}{static case \hspace{0.7cm} }&\multicolumn{3}{c}{rotating case} \\
$B^{1/4}\mathrm{(MeV)}$ & $m_s\mathrm{(MeV)}$&$ M(M_\odot)$ & $R_e$(km) & $M(M_\odot)$ & $R_e$(km) & $\Omega (\mathrm{s}^{-1})$ \\ \hline
120 & 220 &1.9 & 14 & 2.3 & 18 & 6300 \\
120 & 280 & 1.7 & 13 & 2 & 17 & 6300 \\
145 & 0& 2.0 & 11 & 2.4 & 15 & 9400 \\
145& 220 & 1.4 & 10 & 1.7 & 13 & 9300 \\
\end{tabular}
\caption{Rotation properties of maximum mass strange stars for different EOS's.}	\label{table:summary}
\end{center}
\end{table*}


\end{document}